\documentstyle[epsf,12pt]{article}
\newcommand{\be}{\begin{equation}}
\newcommand{\ee}{\end{equation}}
\newcommand{\bear}{\begin{eqnarray}}
\newcommand{\eear}{\end{eqnarray}}
\newcommand{\parno}{\par\noindent}
\def\vec#1{\mathchoice{\mbox{\boldmath$\displaystyle\bf#1$}}
{\mbox{\boldmath$\textstyle\bf#1$}}
{\mbox{\boldmath$\scriptstyle\bf#1$}}
{\mbox{\boldmath$\scriptscriptstyle\bf#1$}}}
\begin{document}
\begin{titlepage}
\noindent
\hfill HD--THEP--96--16 \\
\vspace{3.2cm}
\begin{center}
{\LARGE\bf CP violating couplings}\\
\bigskip
{\LARGE\bf in $Z\rightarrow$ 3 jet decays revisited}\\
\vspace{1cm}
P.\ Haberl\\
\bigskip
Institut  f\"ur Theoretische Physik\\
Universit\"at Heidelberg\\
Philosophenweg 16, D-69120 Heidelberg, Germany\\
\vspace{3.1cm}
{\bf Abstract:}\\[7mm]
\parbox[t]{\textwidth}{
Possible CP violating effects in $Z\rightarrow$ 3 jet decays are 
investigated. The analysis assumes the presence of CP violating 
$Zq\bar qG$ couplings. The contribution of these couplings to the 
$Z\rightarrow q\bar qG$ decay width is calculated for different 
cut algorithms, including nonzero quark masses. Various CP--odd 
observables are discussed and it is shown that their sensitivity
can change significantly if one uses normalized or unnormalized
momentum vectors for their construction. Optimal observables are 
proposed which allow to measure the new couplings simultaneously.}
\end{center}
\end{titlepage}
\section{Introduction}
The large number of $Z$ bosons collected at the LEP collider 
allows detailed studies of $Z$ decays. An interesting possibility
is the search for CP violation beyond the one present in the
Standard Model (SM). For general discussions and lists of further 
work we refer to \cite{blmn,bbbhn}. In this note we want to present 
new calculations for the $Z$ decay rate which have already been 
applied in a recent experimental analysis \cite{aleph}. We further 
reinvestigate certain CP--odd observables and propose the 
measurement of optimal observables.
\par
The theoretical framework is defined in ref.\ \cite{blmn}. There,
possible CP violating effects are parametrized by an effective 
Lagrangian, i.e.\ CP violating formfactors are assumed to be real 
and constant. This effective Lagrangian is being added to the SM and 
calculations are performed at tree level. Including operators of 
dimension 6 and demanding $SU(3)_{\rm C}\times U(1)_{\rm em}$ 
invariance the effective Lagrangian relevant for $Z\rightarrow$ 
3 jet decays contains weak and chromoelectric dipole moments and a 
$Zq\bar qG$ vertex with couplings ${\hat h}_{Vq}$ and ${\hat h}_{Aq}$. 
\par
The couplings ${\hat h}_{Vq/Aq}$ correspond to dimension 6 operators 
which conserve the quark chirality. A possible way to generate such
couplings is through loop diagrams involving non-standard Higgs scalars 
\cite{bbhn}, leading naturally to larger couplings for heavier quarks.
In an analysis of $Z$ decays it is therefore legitimate (albeit
not compelling) to assume that the $b$ quark couplings 
${\hat h}_{Vb/Ab}$ are much larger than the corresponding couplings 
for light quarks. In ref.\ \cite{bbbhn} such a scenario was used to 
explain the gap between the experimental and theoretical values of 
$R_b$ with the existence of large CP violating couplings for the $b$. 
However, new (preliminary) data \cite{blondel} seem to indicate that 
this discrepancy is no longer that significant.
\par
First experimental studies \cite{dhastei} have concentrated on the 
measurement of angular correlations of the jets in 
$Z\rightarrow$ 3 jet decays. In the present 
framework these correlations are only sensitive to the couplings
${\hat h}_{Vq/Aq}$ (the chromoelectric dipole moment does not
contribute and effects due to the weak dipole moment are 
suppressed by a relative factor $m_q/M_Z$). In order to obtain
limits on the couplings ${\hat h}_{Vq/Aq}$ from an independent
measurement, the authors of \cite{aleph} used the differential 
two--jet rate, which requires the knowledge of the contribution of the 
new couplings to the decay rate. To avoid statistical uncertainties 
from event generators, analytic expressions for the rate are useful
and shall therefore be presented in the following section.
\section{Anomalous contributions to the rate}
We consider the decays $Z\rightarrow$ 3 jets on the $Z$ resonance.
At the parton level this is equivalent to the reaction 
\be
e^+e^-\rightarrow Z(p)\rightarrow q(k_-)+\bar q(k_+)+G(k_G)\;,
\label{chain}
\ee
and we assume that the quark flavor (and perhaps also the jet
charge) can be measured.
It is convenient to describe this process in terms of 
production and decay matrices, $\rho$ and $R$ \cite{blmn}.
Taking the $Z$ propagator in Breit--Wigner form and using the
narrow width approximation, the total cross section factorizes,
\be
\sigma(e^+e^-\rightarrow Z\rightarrow q\bar qG)=
\sigma(e^+e^-\rightarrow Z)\frac{1}{\Gamma_Z}
\Gamma(Z\rightarrow q\bar qG)\;,
\ee
with the total $Z$ width $\Gamma_Z$ and the on-shell condition
$p^2=M_Z^2$. The partial width $\Gamma(Z\rightarrow q\bar qG)$ 
is given as integral over the trace term of the decay matrix $R$. 
\par
The phase space for the decay $Z\rightarrow q\bar qG$ can be 
written as
\be
d\phi=\frac{M_Z}{2^8\pi^3}dy_+dy_-
\Theta\Big(y_+y_-(1-y_+-y_-)-r^2(y_++y_-)^2\Big)\Theta({\rm cuts})\;.
\label{phsp}
\ee
The (positive) integration variables $y_\pm$ are defined in terms
of the $\bar q$/$q$ energies $E_\pm$ in the $Z$ rest system as 
$y_\pm=1-2E_\pm/M_Z$; the gluon energy is $E_G=M_Z(y_++y_-)/2$.
We further defined $r=m_q/M_Z$.
The second $\Theta$-function implements the presence of cuts
with cut parameter $y$,
\be
\Theta({\rm cuts})=\Theta(y_{+-}-y)\Theta(y_{+G}-y)\Theta(y_{-G}-y)\;,
\ee
with 
\be
y_{ij}=\left\{\begin{array}{cl}
2M_Z^{-2}E_iE_j(1-\cos\vartheta_{ij}) 
& \mbox{for JADE cuts,} \\[2mm]
2M_Z^{-2}{\rm min}\{E_i^2,E_j^2\}(1-\cos\vartheta_{ij})
& \mbox{for DURHAM cuts.}
\end{array}\right.
\ee
Here $\vartheta_{ij}$ is the angle between the momentum directions 
$\hat{\vec k}_i$ and $\hat{\vec k}_j$ ($i,j=+,-,G$) measured in the 
$Z$ rest system. Note that the JADE algorithm agrees with the so-called 
`E--cuts' $y_{ij}=(k_i+k_j)^2/M_Z^2$ only for massless fermions.
These prescriptions can of course be applied to the decay 
$Z\rightarrow f\bar f\gamma$ as well.
\par
In Fig.~1a) we show the shape of the phase space for massless 
fermions (the dashed triangle) and for $b$ quarks (i.e.\
$r=1/20$), both without cuts. Fig.~1b) shows the effect of JADE and 
DURHAM cuts on the phase space for $b$ quarks. In this case the 
border lines are nontrivial functions given implicitly by the zeroes
of $\Theta({\rm cuts})$ (although they look like straight lines).
\par
The SM result for the rate $\Gamma(Z\rightarrow q \bar qG)$
has been calculated to order $\alpha_s^2$
in \cite{ABinSM}. Since the new couplings $\hat{h}_{Vq/Aq}$
are CP violating, their contribution to the width adds 
incoherently \cite{bbbhn},
\be
\Gamma(Z\rightarrow q\bar qG)=\Gamma^{\rm SM}_{q\bar qG}
+\Delta\Gamma_{q\bar qG}\;.
\ee
Note that in the differential two--jet rate
\be
D_2(y)=\frac{1}{\Gamma(Z\rightarrow q\bar qG)}\frac{d\Gamma(
Z\rightarrow q\bar qG)}{dy}
\ee
the anomalous part is not additive, since also the
normalization is affected. The additional contribution to the 
decay width can be written as
\be
\Delta\Gamma_{q \bar qG}=\frac{\alpha_s}{\pi}\Gamma_{\nu\bar\nu}
\Big(\;f_+(y,r^2)[{\hat h}_{Vq}^2+{\hat h}_{Aq}^2]
+f_-(y,r^2)[{\hat h}_{Vq}^2-{\hat h}_{Aq}^2]\;\Big)
\label{delgam}
\ee
with $\Gamma_{\nu\bar\nu}=\alpha M_Z/(24\sin^2\theta_W\cos^2\theta_W)$.
The coefficient functions $f_\pm(y,r^2)$ are obtained from 
integrating the relevant parts of the trace term $a$ of the 
decay matrix $R$ (cf.\ eq.\ (\ref{secndo})) over the phase space 
volumes shown in Fig.~1b).
Note that we consider only the contributions from the couplings 
${\hat h}_{Vq/Aq}$, i.e.\ we assume the dipole moments to vanish.
Their contributions to the rate can be computed in a similar way
with the full trace term $a$.
\par
The couplings ${\hat h}_{Vq/Aq}$ correspond to a pointlike interaction 
which is not present in the SM Lagrangian. Since there are no internal
propagators, infrared singularities are absent and the anomalous 
contribution $\Delta\Gamma_{q \bar qG}$ is therefore finite even if 
no cuts are applied. In this case one finds
\parno
\hfill\parbox{13.8cm}{
\begin{eqnarray*}
f_+(0,r^2)&=&\frac{2}{5}\Big(1-\frac{41}{2}r^2-29r^4+5r^6
+30r^8\Big)\sqrt{1-4r^2}\\
&+&48r^4\Big(1-r^2+r^6\Big)\ln\left(
\frac{1+\sqrt{1-4r^2}}{2r}\right)\;,\\
f_-(0,r^2)&=&\frac{r^2}{3}\Big(
3+26r^2-62r^4+60r^6\Big)\sqrt{1-4r^2}\\
&+&16r^4\Big(-1+3r^2-6r^4+5r^6\Big)\ln\left(
\frac{1+\sqrt{1-4r^2}}{2r}\right)\;.     
\end{eqnarray*} }
\hfill\parbox{0.7cm}{\begin{eqnarray}  \end{eqnarray} }
\parno
In the presence of cuts the coefficient functions $f_\pm$ can be
computed analytically only for massless fermions. For nonzero
fermion masses one can at least give a series expansion
\be
f_\pm(y,r^2)=f^{(0)}_\pm(y)+r^2f^{(1)}_\pm(y)+O(r^4,r^4\ln r^2)\;,
\label{series}
\ee
where the neglected terms for the heaviest fermion in question
-- the $b$ quark -- are of the order $4\cdot 10^{-5}$.
In (\ref{series}), $f^{(0)}_\pm(y)$ represents the
exact result for massless fermions.
For JADE cuts, the coefficient functions are
\parno
\hfill\parbox{13.8cm}{
\begin{eqnarray*}
f^{(0)}_+(y)&=&\frac{14}{27}(1-3y)^2-\frac{1}{9}(1-3y)^4
-\frac{1}{135}(1-3y)^5\;,\\   
f^{(1)}_+(y)&=&(1-3y)\left(-9-\frac{107}{3}y+13y^2-3y^3\right)
+8y(5-y^2)\ln\left(\frac{1-y}{2y}\right)\;,\\
f^{(0)}_-(y) &=& 0 \;,\\
f^{(1)}_-(y) &=& \frac{8}{9}(1-3y)^2+\frac{1}{9}(1-3y)^4\;.
\end{eqnarray*} }
\hfill\parbox{0.7cm}{\begin{eqnarray}  \end{eqnarray} }
\parno
For DURHAM cuts we obtain
\parno
\hfill\parbox{13.8cm}{
\begin{eqnarray*}
f^{(0)}_+(y)&=&\frac{1}{60}\Big(
24-350y-720y^2+2570y^3+525y^4-2997y^5\Big)\\
&+&\frac{y}{60}\Big(400-754y-321y^2+999y^3\Big)\sqrt{y(8+y)}\\
&+&8y^2\Big(-3+5y+4y^2-6y^3\Big)
\ln\left(\frac{3y+\sqrt{y(8+y)}}{8y}\right)\;,\\
f^{(1)}_+(y)&=&\frac{1}{6}\Big(-54-170y+1104y^2+29y^3+351y^4\Big)\\
&+&\frac{1}{6}\Big(120-380y+5y^2-117y^3\Big)\sqrt{y(8+y)}\\
&+&8y\Big(-13+20y-3y^2+7y^3\Big)\ln\left(
\frac{3y+\sqrt{y(8+y)}}{8y}\right)\;,\\
f^{(0)}_-(y)&=&0\;,\\
f^{(1)}_-(y)&=&\frac{1}{6}\Big(6+28y+132y^2-322y^3+351y^4\Big)\\
&+&\frac{y}{6}\Big(-68+122y-117y^2\Big)\sqrt{y(8+y)}\\
&+&8y\Big(-2+5y-10y^2+7y^3\Big)\ln\left(
\frac{3y+\sqrt{y(8+y)}}{8y}\right)\;.
\end{eqnarray*} }
\hfill\parbox{0.7cm}{\begin{eqnarray}  \end{eqnarray} }
\parno
As a nontrivial check on these results we have
$f^{(0,1)}_\pm(1/3)=0$ since for the cut parameter $y=1/3$ 
the phase space has shrunk to a point.
\par
In Fig.~2 we show the coefficient functions $f_\pm(y,r^2)$
of (\ref{delgam}) for $b$ quarks, i.e.\ for $r=1/20$,
both for the JADE and the DURHAM algorithm.
For massless quarks, $f_+(0,0)=2/5$ whereas $f_-(y,0)$ 
vanishes identically. For nonzero quark masses
the first expansion coefficients in (\ref{series})
are larger than the leading terms by a factor of $\sim$20.
As a consequence, the mass corrections for $b$ quarks amount 
to 5\% (and not -- as expected -- to $r^2\simeq$ 0.25\%).
\par
For practical purposes one can neglect the contribution
proportional to $[{\hat h}_{Vq}^2-{\hat h}_{Aq}^2]$ in (\ref{delgam}).
Assume first that $f_-$ vanishes. An experimental limit
on the size of $\Delta\Gamma$ translates then into
a bound on $({\hat h}_{Vq}^2+{\hat h}_{Aq}^2)^{1/2}$ which means
that allowed couplings lie within a circle with a certain radius 
$R$ in the ${\hat h}_{Vq}$-${\hat h}_{Aq}$-plane. If $f_-$ is 
turned on, this circle is deformed to an ellipse with half
axes $R/\sqrt{1\pm \varepsilon}$ with $\varepsilon=f_-/f_+$.
A numerical study shows however that $0.004\le\varepsilon\le 0.006$
for all values of the cut parameter $y$, both for JADE and
DURHAM cuts.
\section{CP--odd observables}
The best way to measure CP violating couplings is of course with
CP--odd observables. Due to fragmentation effects, the parton spins 
cannot be reconstructed, and therefore CP--odd observables have 
to be built from parton momentum directions (the parton energies 
are determined once the angles are known \cite{aleph}). This is in 
contrast to $t\bar t$ production (see e.g.\ \cite{bra1}) where
the top spin can be traced through the decay.
Observables that use only momentum directions, i.e.\ pure
angular correlations, were proposed and investigated in \cite{blmn}:
\bear
V^{(1)}_{i} &=& \left(\frac{\hat{\vec k}_+\!\times\hat{\vec k}_-}
{|\hat{\vec k}_+ \!\times\hat{\vec k}_-|}\right)_i\;,\label{v1}\\
T^{(1)}_{ij} &=& \left(\hat{\vec k}_+ -\hat{\vec k}_-\right)_i
\left(\frac{\hat{\vec k}_+\!\times\hat{\vec k}_-}
{|\hat{\vec k}_+ \!\times\hat{\vec k}_-|}\right)_j
+(i\leftrightarrow j)\;,\label{t1}
\eear
where $i$, $j$ denote Cartesian indices in the $Z$ rest system.
In addition we consider the following set of CP--odd observables:
\bear
V^{(2)}_{i} &=& (\hat{\vec k}_+\!\times\hat{\vec k}_-)_i\;,\\
V^{(3)}_{i} &=& |\hat{\vec k}_+ \!\times\hat{\vec k}_-|
(\hat{\vec k}_+\!\times\hat{\vec k}_-)_i\;,\\
T^{(2)}_{ij} &=& (\hat{\vec k}_+ -\hat{\vec k}_-)_i
(\hat{\vec k}_+\!\times\hat{\vec k}_-)_j+(i\leftrightarrow j)\;,\\
T^{(3)}_{ij} &=& |\hat{\vec k}_+ \!\times\hat{\vec k}_-|
(\hat{\vec k}_+ -\hat{\vec k}_-)_i(\hat{\vec k}_+\!\times
\hat{\vec k}_-)_j+(i\leftrightarrow j)\;. \label{t3}
\eear
These differ from (\ref{v1}), (\ref{t1}) only by factors 
of $|\hat{\vec k}_+ \!\times\hat{\vec k}_-|=\sin\vartheta_{+-}$.
In the effective Lagrangian approach, the anomalous couplings
$\hat{h}_{Vq/Aq}$ correspond to pointlike interactions, which 
typically lead to event topologies which are not collinear.
An additional weight factor $\sin\vartheta_{+-}$ suppresses
collinear events and should therefore increase the sensitivity 
to $\hat{h}_{Vq/Aq}$.
\par
The observables listed above have definite transformation behaviour 
(as vectors and tensors), thus their expectation values are 
proportional to the $Z$ vector and tensor polarization 
$\vec s$ and $s_{ij}$. For unpolarized beams and taking the
$e^+$ beam as 3-direction we have $\vec{s}=(0,0,\gamma_{\rm el})$
and $s_{ij}=\frac{1}{6}\mbox{diag}(-1,-1,2)$ (here 
$\gamma_{\rm el}=2g_{Ve}g_{Ae}/(g_{Ve}^2+g_{Ae}^2)$ with
the weak vector and axial vector $Ze^+e^-$ couplings $g_{Ve}$
and $g_{Ae}$). Thus what one really should measure are the
components $V^{(n)}_{3}$ and $T^{(n)}_{33}$. Also note that the 
tensor observables $T^{(n)}_{ij}$ do not change sign upon charge 
misidentification
$\hat{\vec k}_+\leftrightarrow\hat{\vec k}_-$, whereas for the 
vector observables the correct assignment of the jet charge 
is required, which makes them less useful experimentally.
\par From 
now on we assume that only the $b$ quarks have nonvanishing
$\hat{h}_{Vq/Aq}$ couplings. The expectation values of the
observables should then be evaluated in an event sample 
containing $Z\rightarrow b\bar bG$ decays
whereas $Z$ decays into light quarks can be used to test the
CP blindness of the detector.
\par
For the number $N_{\rm cut}$ of $Z\rightarrow b\bar bG$ events within
cuts needed to see a 1 s.d.~effect we have (writing generically
${\cal O}$ for one of the observables)
\be
N_{\rm cut}=\frac{\langle{\cal O}^2\rangle}
{\langle{\cal O}\rangle^2}\;.
\ee
Following ref.\ \cite{bbbhn} we scale this number 
\be
N=\frac{\Gamma_Z}{\Gamma(Z\rightarrow b\bar bG)}N_{\rm cut}
\label{scale}
\ee
in order to give an estimate of the {\it total} number $N$ of
$Z$ bosons (note however that this procedure disregards 
detector resolutions, $b$ tagging efficiencies etc.).
\par
For the analysis of CP--odd observables we neglect terms
quadratic in CP violating couplings. The variances 
$\langle{\cal O}^2\rangle$ are then determined from the
SM amplitude, whereas the expectation values are linear
functions of the anomalous couplings. In this approximation the 
sensitivity of the observable $\cal O$ defines a band in the 
$\sqrt{N}\,\hat{h}_{Vb}$--$\sqrt{N}\,\hat{h}_{Ab}$ plane. In the 
case of the tensor observables $T^{(n)}_{ij}$ the slope 
of this band is given by the ratio $g_{Vb}/g_{Ab}$ of weak 
$Zb\bar b$ couplings, since the expectation values 
$\langle T^{(n)}_{ij}\rangle$ are proportional to the combination 
$\hat{h}_{b}=\hat{h}_{Vb}g_{Ab}-\hat{h}_{Ab}g_{Vb}$ \cite{bbbhn}
(the latter remains true also if nonzero dipole moments
are allowed, cf.\ eq.\ (\ref{firsto})).
\par
In Fig.~3 we show the result for the observables
(\ref{v1})--(\ref{t3}). Plotted is the half width of the band
in the $\sqrt{N}\,\hat{h}_{Vb}$--$\sqrt{N}\,\hat{h}_{Ab}$ plane
as function of the cut parameter $y$. If $y$ tends to zero,
this width diverges for the observables $V^{(1)}_{3}$ and 
$T^{(1)}_{33}$ (long dashed curves). 
The best observables are $V^{(2)}_{3}$ and 
$T^{(2)}_{33}$ (solid lines) whereas the sensitivity of 
$V^{(3)}_{3}$ and $T^{(3)}_{33}$ is slightly reduced.
The gain in sensitivity when going from $V^{(1)}_{3}$  
($T^{(1)}_{33}$) to $V^{(2)}_{3}$ ($T^{(2)}_{33}$)
can be easily understood: The unit vector
$\hat{\vec n}=(\hat{\vec k}_+\!\times\hat{\vec k}_-)/
|\hat{\vec k}_+ \!\times\hat{\vec k}_-|$ used in the
construction of the observables $V^{(1)}_{3}$ and $T^{(1)}_{33}$
gives only the direction normal to the decay plane, whereas the 
unnormalized vector $\hat{\vec k}_+\!\times\hat{\vec k}_-$
has an additional information content, namely the angle between 
$\hat{\vec k}_+$ and $\hat{\vec k}_-$. 
\par
As a byproduct we have also recalculated the results of ref.\
\cite{bbbhn} for nonzero $b$ quark mass (in \cite{bbbhn} 
$m_b=0$ was used throughout) and found that the numbers $N$ 
given there increase typically by 3\% -- 8\%.
\section{Optimal observables}
In this section we construct a set of optimal observables to
measure the couplings $\hat{h}_{Vb/Ab}$ separately, following
the procedure developed in \cite{opti1,opti2}. Central assumptions
are that one can neglect higher orders in the anomalous couplings
and that the usual Gaussian error analysis applies.
In linear approximation one writes the differential cross
section for the reaction (\ref{chain}) as
\be
d\sigma(e^+e^-\rightarrow Z\rightarrow b\bar bG)=d\phi^\prime\left(
S_0+S_V\hat{h}_{Vb}+S_A\hat{h}_{Ab}+O(\hat{h}^2_{Vb/Ab})\right)\;,
\ee
where $S_0$ represents the SM contribution. The phase space 
measure $d\phi^\prime$ includes in addition to $d\phi$ of eq.\
(\ref{phsp}) two integrations over angles relative
to the $e^+$ beam direction $\hat{\vec p}_+$.
According to \cite{opti2}, the optimal observables are then 
given by the ratios
\be
{\cal O}_{V/A}=\frac{S_{V/A}}{S_0}\;,
\ee
and any nonzero expectation value $\langle{\cal O}_{V/A}\rangle\neq 0$
signals CP violation. From the explicit formulae one easily
derives
\parno
\hfill\parbox{13.8cm}{
\begin{eqnarray*}
S_0 &=& a^{(0)}+\gamma_{\rm el}\left[(\hat{\vec p}_+\hat{\vec k}_+)
b_1^{(0)}+(\hat{\vec p}_+\hat{\vec k}_-)b_2^{(0)}\right]+\frac{1}{2}
\left[(\hat{\vec p}_+\hat{\vec k}_+)^2-\frac{1}{3}\right]c_1^{(0)}+\\
&& +\frac{1}{2}\left[(\hat{\vec p}_+\hat{\vec k}_-)^2-\frac{1}{3}
\right]c_2^{(0)}+\left[(\hat{\vec p}_+\hat{\vec k}_+)(\hat{\vec p}_+
\hat{\vec k}_-)-\frac{1}{3}(\hat{\vec k}_+\hat{\vec k}_-)
\right]c_3^{(0)} \;,\\
S_{V/A} &=& \pm \beta^2C_FN_c
\frac{2}{M_Z^2}\hat{\vec p}_+({\vec k}_+\!\times
{\vec k}_-)\Bigg\{g_{Ab/Vb}\frac{2}{M_Z}\left[\frac{\hat{\vec p}_+
{\vec k}_+}{y_-}-\frac{\hat{\vec p}_+{\vec k}_-}{y_+}\right]+ \\
&& +g_{Vb/Ab}\gamma_{\rm el}\left[-\frac{y_+}{y_-}-\frac{y_-}{y_+}
+(1\pm4r^2)\left(\frac{1}{y_+}+\frac{1}{y_-}\right)\right]\Bigg\}\;.
\end{eqnarray*} }
\hfill\parbox{0.7cm}{\begin{eqnarray} \label{s0va} \end{eqnarray} }
\parno
The term $S_0$ contains only SM quantities; the relevant contributions
$a^{(0)},\ldots ,c_3^{(0)}$ to the coefficients of the decay 
matrix $R$ can be found in \cite{blmn}. The global factors $\beta$,
$C_F$, and $N_c$ are defined in the Appendix.
\par
The general formula given in \cite{opti2} for the sensitivity of these 
observables simplifies for the case of CP violating couplings. 
Since there are no contributions to the rate linear in 
$\hat{h}_{Vb/Ab}$, we have immediately
$\int\! d\phi^\prime\; S_V=\int\! d\phi^\prime\; S_A=0$.
Moreover, only the trace term $a$ contributes to the rate,
i.e.\ $\int\! d\phi^\prime\; S_0=\int\! d\phi^\prime\; a$.
The sensitivity is then determined by the matrix
\be
{\cal M}_{ij}=\left(\int\! d\phi^\prime\; a\right)^{-1}
\int\! d\phi^\prime\;\frac{S_iS_j}{S_0}\;,
\label{mij}
\ee
which defines -- after scaling according to eq.\ (\ref{scale}) -- 
an ellipse in the 
$\sqrt{N}\,\hat{h}_{Vb}$--$\sqrt{N}\,\hat{h}_{Ab}$ plane. In Fig.~4
we show the results of our numerical evaluation. The strong 
correlation in direction of the line $\hat{h}_b=0$ is not 
surprising, since the other combination of couplings,
$\hat{h}_{Vb}g_{Vb}\!-\!\hat{h}_{Ab}g_{Ab}$, 
is suppressed by a factor of $\gamma_{\rm el}\simeq 0.16$.
In the same plot we compare the results for the best vector
observable $V^{(2)}_3$ (long dashed) and the best tensor
observable $T^{(2)}_{33}$ (short dashed). The optimal observables
have a higher sensitivity, although the difference is only small.
Note however that the construction of the optimal observables
requires a measurement of the jet charge (cf.\ eq.\ (\ref{s0va})),
similar to the vector observables.
\section{Conclusions}
In this paper we have presented various calculations which can be 
applied in a search for CP violation in $Z\rightarrow$ 3 jet decays. 
The effect of CP violating $Zq\bar qG$ couplings on the partial width 
$\Gamma(Z\rightarrow q\bar qG)$ was computed for different
cut algorithms and for nonzero quark masses. 
An important result of our investigation of CP--odd observables
was that the sensitivity of the conventional vector and tensor 
observables changes significantly depending on whether normalized
or unnormalized momentum vectors are used for their construction.
The analysis of optimal observables showed that the gain in
sensitivity is only small. Nevertheless their experimental 
exploration might be useful in order to perform a complete 
analysis of the data.
\par\vspace{1.5cm}
\subsection*{Acknowledgments}
I would like to thank S.~Dhamotharan, J.~v.~Krogh, M.~Steiert, 
and M.~Wunsch for their continuous interest in this work and
in particular H.~Stenzel for discussions which lead to part 
of this project. Special thanks are due to O.~Nachtmann for 
valuable discussions and a critical reading of the manuscript.
\par\vspace{1.5cm}
\section*{Appendix}
\renewcommand{\theequation}{A.\arabic{equation}}
\setcounter{equation}{0}
The decay matrix $R$ for the process $Z\rightarrow q\bar qG$
is defined as 
\be
R_{ij}=\sum\langle Z(\vec{e}_i)|
{\cal T}^{\hspace*{1pt}\dagger}|q\bar qG \rangle\;
\langle q\bar qG |{\cal T}|Z(\vec{e}_j)\rangle\;,
\ee
where the summation extends over the $q$ and $\bar q$ spins and 
the gluon polarisation. We work in the $Z$ rest system, where 
$Z(\vec{e}_i)$ denotes a $Z$ with polarization vector along
the $i^{\rm th}$ Cartesian coordinate.
The decomposition of $R$ with the help of the $q$ and $\bar q$
momentum unit vectors $\hat{\vec k}_-$ and $\hat{\vec k}_+$ is  
given in \cite{blmn}. Note however that we did not normalize $R$ 
with $\Gamma_{q\bar qG}$ and therefore we have
$\frac{1}{3}\int\!d\phi\,{\rm Tr}(R)=\Gamma_{q\bar qG}$ (instead of 
$\frac{1}{3}\int\!d\phi\,{\rm Tr}(R)=1$ in \cite{blmn}).
\par
The coefficients of that decomposition are calculated at
tree level using in addition to the SM Lagrangian the
effective CP violating Lagrangian of \cite{blmn} which 
contains weak and chromoelectric dipole moments $\hat{\tilde{d_q}}$
and $\hat{d^\prime_q}$ and the $Zq\bar qG$ couplings
$\hat{h}_{Vq}$ and $\hat{h}_{Aq}$. Contributions from the SM and 
terms linear in CP violating couplings have already been given 
in \cite{blmn}. In order to make this article self-contained 
and to illustrate the different notations, we include the
latter here as well:
\par\noindent
\hfill\parbox{13.8cm}{
\begin{eqnarray*}
a^{(1)}(y_+,y_-) \!&=&\! b^{(1)}_1(y_+,y_-) 
  \;=\; b^{(1)}_2(y_+,y_-) \;=\; 0 \;,\\
b^{(1)}_3(y_+,y_-) \!&=&\! \frac{w(y_+,y_-)}{y_+y_-}\Bigg[
   (\hat{h}_{Vq}g_{Vq}\!-\!\hat{h}_{Aq}g_{Aq}) 
   \Big(y_+(1-y_+)+y_-(1-y_-)\Big)+ \\
&& \! + \; 4\Big( r\hat{\tilde{d_q}} g_{Aq} +r^2 (\hat{h}_{Vq}g_{Vq}
   \!+\!\hat{h}_{Aq}g_{Aq})\Big) (y_++y_-) \Bigg] \;,\\
c^{(1)}_1(y_+,y_-) \!&=&\!  c^{(1)}_2(y_+,y_-) 
  \;=\; c^{(1)}_3(y_+,y_-) \;=\; 0 \;,\\
c^{(1)}_4(y_+,y_-) \!&=&\! -c^{(1)}_5(y_-,y_+) \;=\; 
   \hat{h}_{q}\frac{w(y_+,y_-)}{y_-} \sqrt{(1-y_+)^2-4r^2} \;.
\end{eqnarray*} }
\hfill\parbox{0.7cm}{\begin{eqnarray} \label{firsto} \end{eqnarray} }
\par\noindent
where
\par\noindent
\hfill\parbox{13.8cm}{
\begin{eqnarray*}
\hat{h}_{q} &=& \hat{h}_{Vq}g_{Aq}\!-\!\hat{h}_{Aq}g_{Vq} \;, \\
w(y_+,y_-) &=& \sqrt{y_+y_-(1-y_+-y_-)-r^2(y_++y_-)^2} \;=\;
   2\frac{|{\vec k}_+\!\times{\vec k}_-|}{M_Z^2} \;.
\end{eqnarray*} }
\hfill\parbox{0.7cm}{\begin{eqnarray} \label{hhat} \end{eqnarray} }
\par\noindent
Here $g_{Vq}=T_3-2Q_q\sin^2\theta_W$ and $g_{Aq}=T_3$ are the 
SM $Zq\bar q$ couplings; for the definition of $y_\pm$ and $r$ 
see eq.~(\ref{phsp}).
Out of the contributions quadratic in CP violating couplings
we list only the trace term $a^{(2)}(y_+,y_-)$ and the vanishing 
coefficients:
\par\noindent
\hfill\parbox{13.8cm}{
\begin{eqnarray*}
a^{(2)}(y_+\!\!\!\!\!\!&,&\!\!\!\!\!y_-)= \frac{8}{3}\Bigg\{
    \hat{\tilde{d_q}}^2 \Bigg[ \frac{(y_++y_-)^2}{2y_+y_-}
    - 2(y_+ + y_-) + (1-4r^2)\frac{1-y_+-y_-}{y_+y_-}  \\
&-&\!\!r^2(1- 4r^2)\frac{(y_++y_-)^2}{(y_+y_-)^2} \Bigg] \!
    +\, \hat{\tilde{d_q}}\hat{d^\prime_q}g_{Vq} \left[3-2y_+ -2y_- 
    - r^2\frac{(y_++y_-)^2}{y_+y_-} \right] \\
&+&\!\! \hat{\tilde{d_q}}\hat{h}_{Aq}\left[
    - \,\frac{r}{2}\frac{(y_++y_-)^3}{y_+y_-} + 2r(1+y_++y_-)
    - 2r^3\frac{(y_++y_-)^2}{y_+y_-} \right] \\
&+&\!\! \hat{d^\prime_q}^2g_{Aq}^2\left[ \frac{1}{2}y_+y_- 
    + r^2( -3+y_+ +y_- )+r^4\frac{(y_++y_-)^2}{y_+y_-} \right] \\
&+&\!\! \hat{d^\prime_q}^2(g_{Vq}^2+g_{Aq}^2)\frac{1}{2}(1-y_+-y_-)
    \;+\; \hat{d^\prime_q}\hat{h}_{Aq}g_{Vq} r(y_++y_-)  \\
&+&\!\! \hat{d^\prime_q}\hat{h}_{Vq}g_{Aq}\left[\frac{r}{2}
    (y_++y_-) - \frac{r^3}{2} \frac{(y_++y_-)^3}{y_+y_-} \right]  
    +(\hat{h}_{Vq}^2\!-\!\hat{h}_{Aq}^2)\frac{r^2}{4}(y_++y_-)^2 \\
&+&\!\! (\hat{h}_{Vq}^2\!+\!\hat{h}_{Aq}^2)\frac{1}{8}\Big( 
    4y_+y_- +(y_+^2+y_-^2)(1-y_+-y_-) \Big) \Bigg\} \;,\\
b^{(2)}_3(y_+\!\!\!\!\!\!&,&\!\!\!\!\!y_-)=
    c^{(2)}_4(y_+,y_-) \;=\; c^{(2)}_5(y_+,y_-) \;=\; 0 \;.
\end{eqnarray*} }
\hfill\parbox{0.7cm}{\begin{eqnarray} \label{secndo} \end{eqnarray} }
\par\noindent
The nonvanishing coefficients $b^{(2)}_1$, $b^{(2)}_2$, $c^{(2)}_1$, 
$c^{(2)}_2$, and $c^{(2)}_3$ which do not contribute to the rate
but are relevant for other (CP--even) observables have been calculated 
as well and are available on request.
\par
All coefficients in (\ref{firsto}), (\ref{secndo}) have to be 
multiplied with the global factor $\beta^2C_FN_c$, with the coupling  
$\beta=eg_s/(\sin\theta_W\cos\theta_W)$ and the color factor 
$C_F\cdot N_c=\frac{4}{3}\cdot 3$. 
The coefficients can of course be applied to the process
$Z\rightarrow f\bar f\gamma$ as well. In this case one has to
replace the chromoelectric dipole moment $\hat{d^\prime_q}$
with the electric dipole moment $-\hat{d}_f$ and the $Zq\bar qG$
couplings $\hat{h}_{Vq/Aq}$ with the $Zf\bar f\gamma$ couplings 
$-\hat{f}_{Vf/Af}$ (the minus signs stem from the normalization
in \cite{blmn}). Moreover, the global factor is obtained from replacing
in $\beta$ the strong coupling constant $g_s$ with $eQ_f$ and
changing the color factor $C_F\!\cdot\!N_c$ to $1\!\cdot\!1$ 
($1\!\cdot\!3$) for leptons (quarks).
\par\vspace{1.5cm}
\par\vspace{3cm}
\section*{Figure Captions}
\begin{description}
\item[Figure 1:] \quad Allowed phase space regions (Dalitz plots) 
     for the decays $Z\rightarrow f\bar fV$ ($V=$ photon or gluon)
     in the $y_+$--$y_-$ plane. 
     {\bf a)} Without cuts, for massless fermions (dashed)
     and for $b$ quarks (solid).
     {\bf b)} With cut $y=0.1$, for $b$ quarks with JADE (solid)
     and DURHAM algorithm (dashed).
\item[Figure 2:] \quad The coefficients $f_+(y,r^2)$ and $f_-(y,r^2)$
     in the decomposition (\ref{delgam}) shown as function of the
     cut parameter $y$ for JADE (solid) and DURHAM cuts (dashed).
     The parameter $r=m_q/M_Z$ is put to $1/20$ ($b$ quark). 
\item[Figure 3:] \quad {\bf a)} The half width of the sensitivity 
     band in the $\sqrt{N}\,\hat{h}_{Vb}$--$\sqrt{N}\,\hat{h}_{Ab}$ 
     plane as function of the cut parameter $y$. Shown are the 
     results for the vector observables $V_3^{(1)}$ (long dashed),
     $V_3^{(2)}$ (solid) and $V_3^{(3)}$ (short dashed).
     {\bf b)} The same for the tensor observables $T_{33}^{(1)}$ 
     (long dashed), $T_{33}^{(2)}$ (solid) and $T_{33}^{(3)}$ 
     (short dashed).
\item[Figure 4:] \quad Sensitivity contours of different observables 
     in the $\sqrt{N}\,\hat{h}_{Vb}$--$\sqrt{N}\,\hat{h}_{Ab}$ 
     plane. The results for the best vector observable 
     $V_3^{(2)}$ (long dashed) and the best tensor observable 
     $T_{33}^{(2)}$ (short dashed) are compared to the ellipse
     from the optimal observables ${\cal O}_V$, ${\cal O}_A$
     (solid).
\end{description}
\newpage
%
%
\unitlength1.0cm
\noindent
\begin{picture}(15.,8.5)
\put(0.,8.){\makebox(15.,1.){\bf Figure 1}}
\put( 3.5,6.){\makebox(4.,1.){\bf a) }}
\put(11.,6.){\makebox(4.,1.){\bf b) }}
\put(-2.5,-11.8){ \epsfysize=22.cm \epsffile{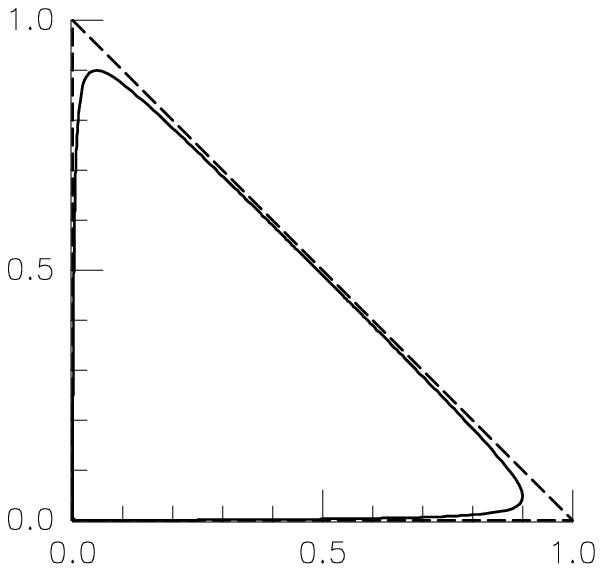} }
\put( 5.,-11.8){ \epsfysize=22.cm \epsffile{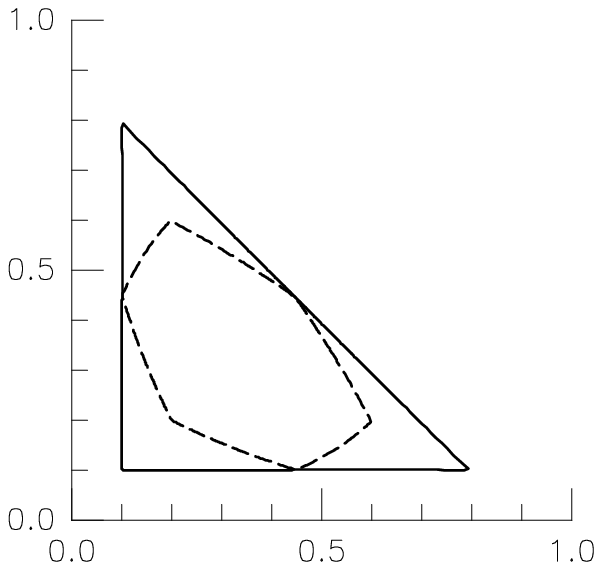} }
\end{picture}
%
\parno
%
\unitlength1.0cm
\begin{picture}(15.,12.)
\put(0.,11.2){\makebox(15.,1.){\bf Figure 2}}
\put( 9.,9.){\makebox(2.,1.){ $f_+(y,r^2)$ }}
\put( 9.,3.9){\makebox(2.,1.){ $f_-(y,r^2)$ }}
\put(11.,0.){\makebox(2.,1.){ $y$ }}
\put(0.1,-6.6){ \epsfysize=22.cm \epsffile{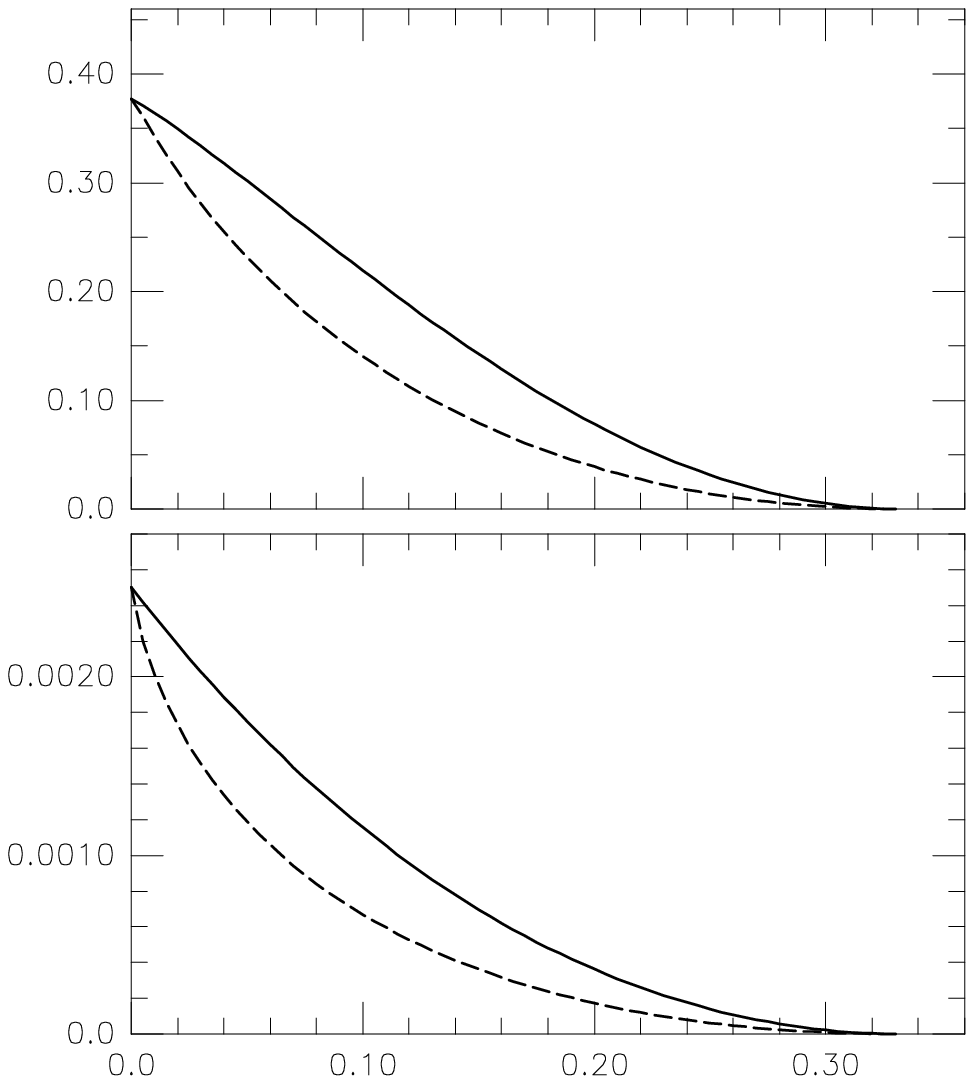} }
\end{picture}
%
\newpage
%
\unitlength1.0cm
\noindent
\begin{picture}(15.,8.5)
\put(0.,8.){\makebox(15.,1.){\bf Figure 3}}
\put( 3.3,5.7){\makebox(4.,1.){\bf a) }}
\put(10.8,5.7){\makebox(4.,1.){\bf b) }}
\put(13.6,0.6){\makebox(2.,1.){ $y$ }}
\put(6.1,0.6){\makebox(2.,1.){ $y$ }}
\put(-3.1,-12.){ \epsfysize=22.cm \epsffile{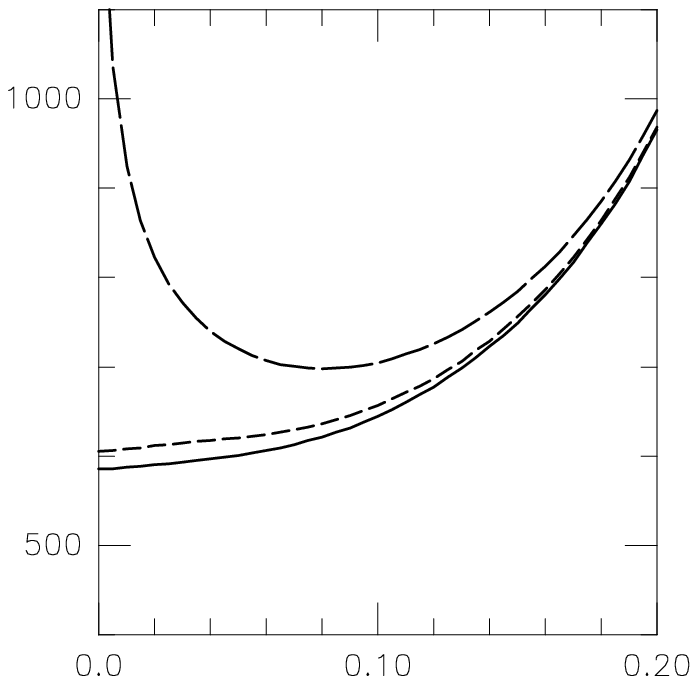} }
\put( 4.4,-12.){ \epsfysize=22.cm \epsffile{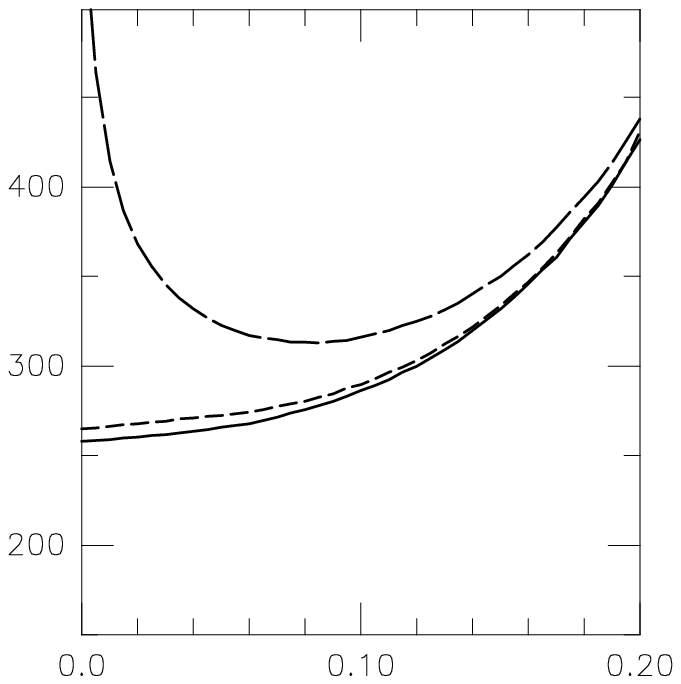} }
\end{picture}
%
\parno
%
\unitlength1.0cm
\begin{picture}(15.,12.)
\put(0.,10.5){\makebox(15.,1.){\bf Figure 4}}
\put( 1.1,9.7){\makebox(2.,1.){ $\sqrt{N}\;\hat{h}_{Vb}$ }}
\put( 12.6,0.6){\makebox(2.,1.){ $\sqrt{N}\;\hat{h}_{Ab}$ }}
\put(-0.8,-4.6){ \epsfysize=22.cm \epsffile{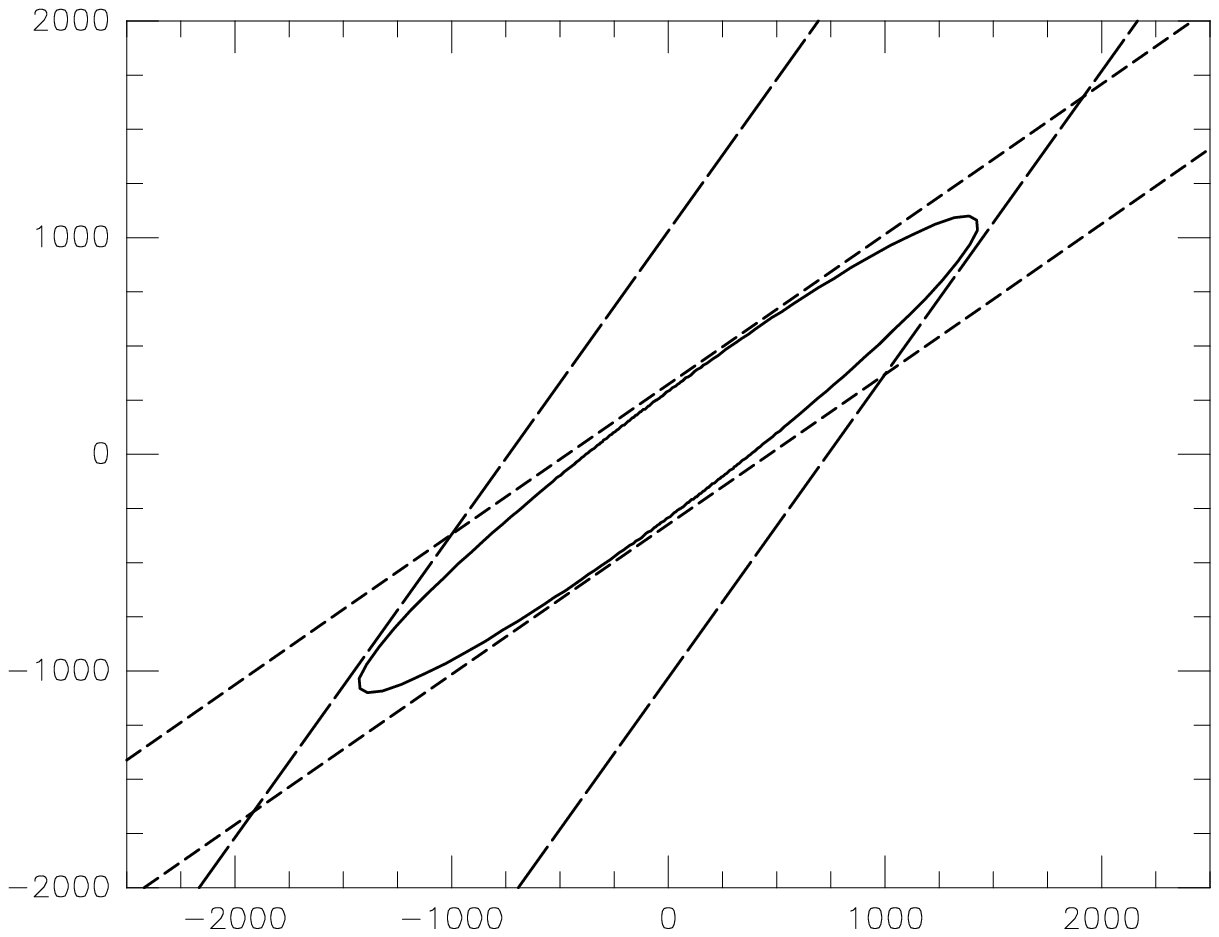} }
\end{picture}
%

\begin{thebibliography}{99}
\bibitem{blmn} W.~Bernreuther, U.~L\"ow, J.~P.~Ma, O.~Nachtmann:
   Z.\ Phys.\ {\bf C 43} (1989) 117
\bibitem{bbbhn} W.~Bernreuther, G.~W.~Botz, D.~Bru\ss, P.~Haberl, 
   O.~Nachtmann: \\ Z.\ Phys.\ {\bf C 68} (1995) 73
\bibitem{aleph} D.~Busculic et al.\ (ALEPH coll.):
   Phys.\ Lett.\ {\bf B 384} (1996) 365
\bibitem{bbhn} W.~Bernreuther, A.~Brandenburg, P.~Haberl, O.~Nachtmann:\\
   Phys.\ Lett.\ {\bf B 387} (1996) 155
\bibitem{blondel} A.~Blondel for the LEP Electroweak Working Group:
   Plenary talk given at the $28^{\rm th}$ Int.~Conf.~on
   High Energy Physics, Warsaw (July 1996)
\bibitem{dhastei} S.~Dhamotharan, Diploma Thesis, University of 
   Heidelberg (unpublished); \\  M.~Steiert, Diploma Thesis, 
   University of Heidelberg (unpublished)
\bibitem{ABinSM} Z.~Kunszt et al.: in `$Z$ physics at LEP1',
   ed.\ G.~Altarelli, CERN Yellow Book 89-08 (1989) 373
\bibitem{bra1} A.~Brandenburg: Phys.\ Lett.\ {\bf B 388} (1996) 626
\bibitem{opti1} D.~Atwood, A.~Soni: Phys.\ Rev.\ {\bf D 45} 
   (1992) 2405;\\ M.~Davier, L.~Duflot, F.~Le Diberder, 
   A.~Roug\'e: Phys.\ Lett.\ {\bf B 306} (1993) 411
\bibitem{opti2} M.~Diehl, O.~Nachtmann: Z.\ Phys.\ {\bf C 62} (1994) 397
\end{thebibliography}
\end{document}